\documentclass[vecphys]{svmult}

\usepackage{graphicx}        
\usepackage[bottom]{footmisc}

\begin{document}

\title*{\textit{Spitzer's} View of Planetary Nebulae}
\author{Joseph L. Hora}
\institute{Harvard-Smithsonian Center for Astrophysics, 60 Garden St.
MS-65, Cambridge, MA 02138 USA;
\texttt{jhora@cfa.harvard.edu}}
%
%
\maketitle
\begin{abstract}

The \textit{Spitzer Space Telescope}, NASA's Great Observatory for infrared
astronomy, has made available new tools for the investigation of the
infrared properties of planetary nebulae.  The three instruments onboard,
including the Infrared Array Camera (IRAC), the Multiband Imaging
Photometer for Spitzer (MIPS), and the Infrared Spectrograph (IRS), provide
imaging capability from 3.6 to 160 $\mu$m, and low and moderate resolution
spectroscopy from 5.2 to 38 $\mu$m.  In this paper I review recent
\textit{Spitzer} results concerning planetary nebulae and their
asymmetrical structures.


\end{abstract}
\section{The \textit{Spitzer} Mission}
\label{sec:1}

The \textit{Spitzer Space Telescope}\ \cite{werner04} (\textit{Spitzer}),
launched in August 2003, is the final component of NASA's ``Great
Observatories'', and its instruments take advantage of the gains in
infrared detector technology since the \textit{IRAS} and \textit{ISO}
missions to achieve unprecedented sensitivity at infrared wavelengths not
available from ground-based observatories.  The Infrared Array Camera
\cite{fazio04}\ (IRAC) obtains images in four bands at 3.6, 4.5, 5.8, and
8.0 $\mu$m.  The Infrared Spectrograph \cite{houck04}\ (IRS) obtains low or
moderate resolution spectra over the range of 5.2 -- 38 $\mu$m.  The
Multiband Imaging Photometer for \textit{Spitzer} \cite{rieke04}\ (MIPS)
obtains images at 24, 70, and 160 $\mu$m as well as a spectral energy
distribution (SED) between 50 and 100 $\mu$m. The telescope is cooled
passively and by helium boil-off vapor to 5.5K, providing a low-background
system that is limited only by the detector properties and the astronomical
background. \textit{Spitzer} allows us to build on the groundbreaking work
done by \textit{IRAS} and \textit{ISO}, providing improved spatial
resolution and a $>$100+ fold increase in sensitivity.

\textit{Spitzer} is an observatory for the community -- most of the time is
awarded to General Observer (GO) programs, including ``Legacy'' projects
which have no proprietary time.  Most of the \textit{Spitzer} data,
including the original guaranteed time (GTO) programs and observations from
the first three GO cycles, are now publicly available.  There are also
further opportunities for submitting new observing proposals.  The next
call for proposals (Cycle 5) will happen in mid-August 2007, with a
proposal deadline of November 16, 2007.  This will be the last proposal
call for the cryogenic mission. The helium cryogen is anticipated to last
until March 2009.  After the cryogen has been depleted, the telescope and
instruments are expected to reach an equilibrium temperature of $\sim$30K,
which will allow the IRAC 3.6 and 4.5 $\mu$m bands to continue to operate
for several more years.  Planning is now underway for a ``warm mission''
which would have an open call for proposals in the fall of 2008.

\section{Application of IR Imaging and Spectroscopy to PNe}
\label{sec:2}

Several examples of the importance of IR imaging and spectroscopy to the
study of planetary nebulae (PNe) were presented at last year's IAU
Symposium \#234 in reviews of IR imaging \cite{hora06a} and spectroscopy
\cite{salas06a}.  Most of those examples apply to the \textit{Spitzer}
data, including the ability to penetrate into optically obscured regions to
resolve structures in the nebula and near the central stars; to measure the
emission from dust and small grains and determine their composition,
properties, and explore aspects of dust formation; to detect emission from
neutral material such as H$_2$ and determine the conditions responsible for
its excitation; and to observe lines of different ionization states in the
IR to determine abundances and physical conditions in the nebulae.

The higher sensitivity of the \textit{Spitzer} instruments allow for the
first time a large sample of Galactic and extragalactic PNe to be studied
in the mid-infrared.  The $5' \times 5'$ field of IRAC and MIPS allow for
relatively rapid imaging of large fields at high sensitivity to map the
distribution of warm and cool dust throughout the nebula and halos of PNe.
The IRS can make pointed observations and also has a spectral mapping mode
which can be used to produce images in various lines and dust features to
determine their spatial distribution.

\section{Available \textit{Spitzer} data}
\label{sec:3}

\textit{Spitzer} is now in its fifth year of operation, and a majority of
the data obtained is now public. The easiest way to find out whether a
source has been observed and what data are available is to use the Leopard
archive query program available from the SSC
website\footnote{http://ssc.spitzer.caltech.edu}.  A list of all approved
programs is also provided, which can be used to search for projects
associated with PNe.

There are several Legacy, GTO, and GO programs that have relevance to PNe.
The GLIMPSE \cite{benj03} and MIPSGAL surveys cover the inner $\pm 65\deg$
of the Galactic disk with IRAC and MIPS bands, respectively.  The GLIMPSE
data have already been used to find new PNe \cite{cohen05} and study
previously known PNe \cite{kwok07,cohen07}.  The SAGE survey imaged the LMC
with IRAC and MIPS at sufficient sensitivity to detect most of the PNe in
that galaxy\cite{hora07}. An IRAC and MIPS survey of the star-forming body
of the SMC was performed early in the mission (S$^3$MC \cite{bolatto07}),
and the approved SAGE-SMC program will map a wider field to map the entire
SMC with IRAC \& MIPS at a similar sensitivity to the LMC.  The approved
SAGE-SPEC Legacy project will survey over 100 objects found in the SAGE
survey, including 12 PNe, 7 Post-AGB, and 72 AGB stars \& candidates.

There are several GTO programs that have focused on PNe.  A series of IRAC
GTO programs have imaged over fifty PNe \cite{hora04} (program IDs (PIDs)
68, 30285, 40020).  Another 16 PNe with dual dust chemistry are to be
observed in Cycle 4 (PID 40115). Several other IRAC programs study young
PNe, Luminous Blue Variables, and PNe halos.  The MIPS GTO program (PID 77)
imaged 10 PNe with MIPS (e.g., \cite{su04}), and obtained IRS spectra of
several objects.  There are IRS GTO programs that study PNe in the LMC
\cite{salas04} and SMC, and in the Galactic Bulge (PIDs 103, 30482, 30550).

In a quick check of the list of approved programs before this conference, I
found 14 GO programs directly associated with PNe, and there are no doubt
more that contain data useful to the study of PNe. Most of the GO programs
use the IRS to obtain spectra to study abundances, dust properties and
evolution, and metallicity effects. There are also programs that use MIPS
to examine the cool dust in PNe.

\section{IRS Results}
\label{sec:4}

A discussion of the IRS results since launch was included in the review of
IR spectroscopy of PNe by Bernard-Salas \cite{salas06a} at IAU Symposium
\#234.  The first published IRS spectrum of a PN was of SMP83 in the LMC
\cite{salas04}, which demonstrated the instrument sensitivity and the
capability to determine abundances in PNe as distant as in this neighboring
galaxy.  Another observation of a LMC PN from the IRS GTO program was of
SMP LMC 11 \cite{salas06}. The optical properties of this object resemble
that of a PN, but its spectrum implies a pre-planetary nebula (PPN).  The
spectrum shows no polycyclic aromatic hydrocarbon (PAH) emission but many
molecular absorption bands, and the first detection of C$_4$H$_2$,
C$_6$H$_2$, and C$_6$H$_6$ absorption in an extragalactic object.  These
molecules are the building blocks from which more complex hydrocarbons are
produced. The spectrum is similar to that seen by \textit{ISO} in the PPN
AFGL 618 \cite{cernicharo01}. SMP LMC 11 shows a lack of nitrogen-based
molecules (apparently a trend in LMC/SMC objects) which are prominent in
Galactic objects.

The post-AGB star MSX SMC 029 has also been observed with IRS
\cite{kraemer06}. The spectrum is dominated by a cool dust continuum
($\sim$280K).  Both PAH emission and absorption from C$_2$H$_2$ are
present. Strong absorption features from C$_2$H$_2$, C$_4$H$_2$, HC$_3$N,
and C$_6$H$_6$ in the 13 -- 16 $\mu$m range are also seen, similar to those
in the post-AGB objects AFGL 618 and SMP LMC 11. The PAH features in the
spectrum are unusual, with a peak emission in the 7 -- 9 $\mu$m complex
beyond 8 $\mu$m instead of near 7.7 -- 7.9 $\mu$m. Also, the 8.6 $\mu$m
feature is as strong as the C -- C mode feature. The 11.3/8 $\mu$m ratio
suggests that the PAHs either have a low-ionization fraction or are
unprocessed, which implies that MSX SMC 029 has only recently evolved off
the AGB branch.

Another investigation of the PNe in the Magellanic Clouds has been
performed which examined a total of 16 SMC and 25 LMC PNe with the IRS
\cite{stang06,stang07}.  They report that about half of the PNe show either
C-rich or O-rich dust compounds. They also find that all PNe with
carbonaceous dust are either round or elliptical while all PNe with O-rich
dust are bipolar. The IRS spectra of symmetric and asymmetric PNe are
apparently extremely different, showing a tight connection between dust
type and morphology.

\section{MIPS Results}
\label{sec:5}

The MIPS observations of PNe have been able to trace the distribution of
the cool dust, as well as forbidden line emission from the ionized gas.
In the first MIPS results reported on NGC 2346 \cite{su04}, the emission is
seen to be strongest in the equatorial plane and the walls of the bipolar
lobes of the nebula, and is in general more extended than the optical
emission.  The data can be fitted using a model with two components with
temperatures of 60K and 25K.  A central peak is seen in the 24 $\mu$m
image, which is likely the dust causing the deep fading of the central star
in the optical. Polar tips are seen in the bipolar lobes which indicate a
more collimated component of the mass loss. A toroidal structure is seen in
the equatorial plane at 70 $\mu$m. The angular offset and the color
difference in the torus between 24 and 70 $\mu$m indicates the PN is not
completely axisymmetric and there has been some change in position angle,
which could have been caused by the binary companion. The PN NGC 650 was
also observed with MIPS \cite{ueta06}, and it was found that the 24 $\mu$m
emission is largely due to the [O {\sc iv}] line at 25.9 $\mu$m, whereas
the 70 and 160 $\mu$m flux is due to $\sim$30K dust continuum emission in
the remnant AGB shell. The far-IR nebula structure suggests that the
enhancement of mass loss at the end of the AGB phase has occurred
isotropically, but has ensued only in the equatorial directions while
ceasing in the polar directions.

The MIPS observations of the Helix found an unresolved central source with
excess thermal continuum emission at 24 and 70 $\mu$m, as well as in the
IRAC 8 $\mu$m band \cite{su07}.  They conclude that the emission is most
likely from a dust disk with temperatures of 90 -- 130K, distributed from
35 -- 150 AU from the central star.  They speculate that this dust possibly
arises from collisions of Kuiper Belt-like objects, or the breakup of
comets from an Oort-like cloud that has survived the post-main sequence
evolution of the star.  The Helix also has an inner bright bubble at 24
$\mu$m due to [O {\sc iv}] which fills in the region inside the main ring.
The 160 $\mu$m emission is primarily in the main ring, with an appearance
similar to that at 8 $\mu$m.

\section{IRAC Results}
\label{sec:6}

\begin{figure}
\centering
\includegraphics[height=10cm]{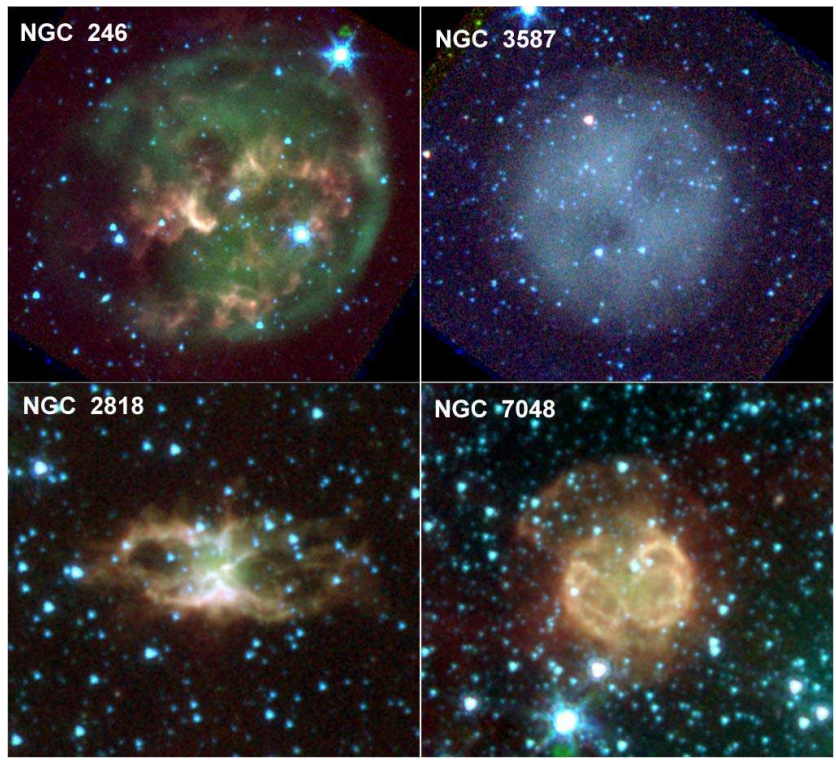}
\caption{IRAC color images of PNe NGC 246, NGC 2818, NGC 3587, and NGC 7048
\cite{hora07a}.  In all images the 3.6, 4.5, and 8.0 $\mu$m bands are
assigned to blue, green, and red, respectively. North is up and East is to
the left in all images.}
\label{fig:1}       
\end{figure}

The first IRAC images of PNe \cite{hora04,hora06b} showed that
\textit{Spitzer} could probe the faint extended emission from ionized gas,
warm dust, PAHs, and H$_2$ in relatively short integration times.  Some
typical results are shown in Figure~\ref{fig:1}. The appearance of the
extended emission in the IRAC bands is similar to the optical appearance in
some cases, but often there are important differences.  For example, in NGC
246 an unexpected ``ring'' of emission is prominent in the longer IRAC
wavelengths within the elliptical shell of the nebula \cite{hora04}.

\begin{figure}
\centering
\includegraphics[height=12cm]{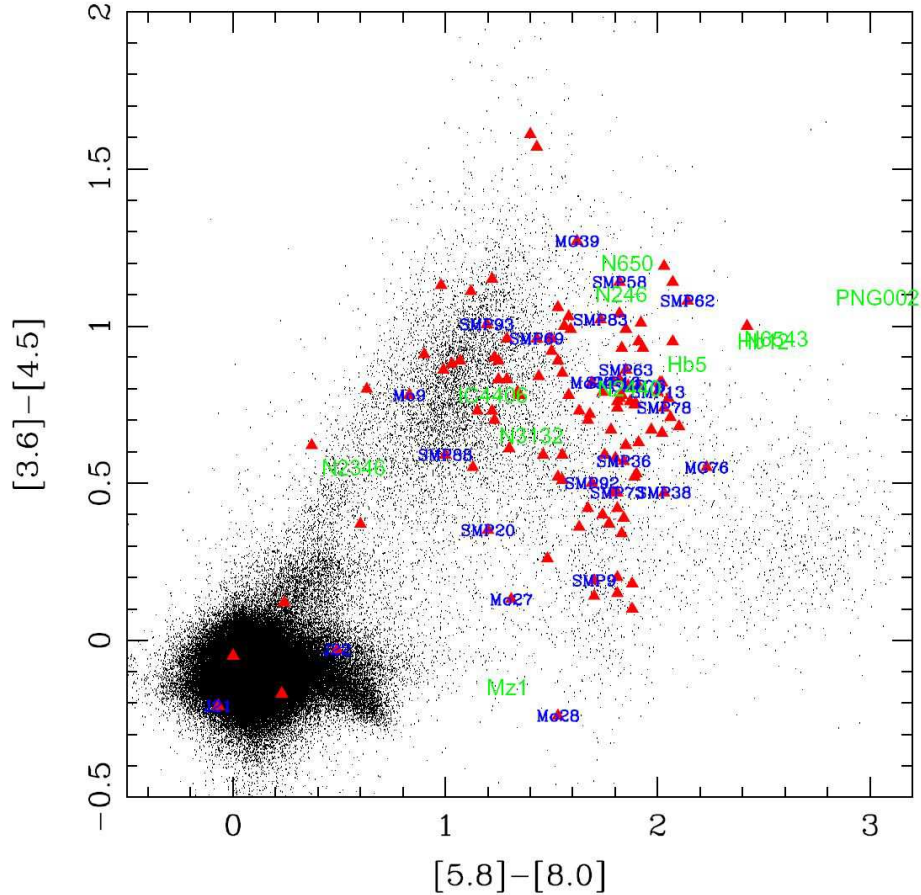}
\caption{The [3.6]-[4.5] versus [5.8]-[8.0] color-color diagram for the LMC
PNe and SAGE catalog sources. The LMC sample \cite{hora07} is plotted as
red triangles, and a subset of these are labeled with blue letters. The
subset was chosen to contain examples of different mid-IR spectral types
\cite{salas05} and optically-determined morphologies \cite{shaw06}. Also
plotted in the IRAC-only diagrams are data from several Galactic PNe that
have been observed with IRAC \cite{hora06a}, shown as the green labels. The
underlying black points are a subsample of the SAGE database detected in
all four IRAC bands.}
\label{fig:2}       
\end{figure}

In PNe that are dominated by H$_2$ emission, such as in NGC 6720, NGC 6853,
and NGC 7293, the spatial distribution closely matches that of the
2.12~$\mu$m H$_2$ line emission \cite{hora06b}.  Because the stellar and
nebular free-free continuum is much reduced at the IRAC wavelengths, the
emission from the halo and from the dust and molecular lines appear more
prominent.  The IRAC colors of PNe are red, especially in the 5.8 and 8.0
$\mu$m bands \cite{hora06b,hora07,kwok07} (see Figure~\ref{fig:2}), which
separates them from main sequence and evolved stars, although they occupy a
region of color-color space similar to young stellar objects and H {\sc ii}
regions \cite{whitney07}.

The GLIMPSE survey has yielded several results on PNe.  The discovery of
the new PN G313.3+00.3 has been reported based on the IRAC imaging and
radio images which show the round PN shell \cite{cohen05}.  The ability to
see through the extinction in the plane, and the higher IRAC resolution
compared to previous IR surveys were important in identifying the new PN.
The positions of known PNe in the survey have been examined, and roughly
one third have been detected \cite{kwok06,kwok07}, even at a relatively
shallow survey depth.  Another set of objects with PNG classifications were
found unlikely to be PNe, based on their IR morphology.

The known PNe in the LMC from the SAGE survey have also been analyzed
\cite{hora07} (see Figure~\ref{fig:2}).  Approximately 75\% of the
optically-identified PNe were detected by IRAC and MIPS at the survey
depth.  Their IRAC colors were found to be similar to the set of Galactic
PNe, and to depend on the dominant spectral characteristics of their mid-IR
spectra as determined from their IRS spectra \cite{salas05}.  The IR PNe
luminosity function (LF) for the LMC was found to follow the same
functional form as the well-established [O {\sc iii}] LF, although there
are several PNe with observed IR magnitudes brighter than the cut-offs in
these LFs.

The IRAC images of the nearby PN NGC 7293 (The Helix) \cite{hora06} resolve
the famous cometary knots previously detected in optical images.  The IRAC
images are dominated by H$_2$ emission (as was first seen by \textit{ISO}
\cite{cox98}), although the excitation mechanism is not clear \cite{hora06,
odell07, matsuura07}.  In addition to the IR emission from the tips of the
knots which are also bright in the optical, H$_2$ emission is detected from
the tails of the knots in the IRAC images.  The main rings are composed of
a large number of clumps similar to the cometary knots but without the long
tails, and radial rays of emission extend from the outer edge of the main
ring into the outer halo.

\section{Summary}
\label{sec:7}

\textit{Spitzer} infrared observations of PNe are providing unique insights
on the composition, ionization, structure, temperature, and relative
location of ionized gas, dust, PAHs, and H$_2$ in the nebulae.  A wealth of
data exists in the \textit{Spitzer} archive now, in large-area surveys and
pointed observations of PNe.  The final Call for Proposals for the
cryogenic mission has been issued and they are due in November 2007.  Act
now if you have a need for new \textit{Spitzer} observations!

\acknowledgement

This work is based in part on observations made with the Spitzer Space
Telescope, which is operated by the Jet Propulsion Laboratory, California
Institute of Technology under a contract with NASA. Support for this work
was provided by NASA through an award issued by JPL/Caltech.

%
%



\end{document}